 \documentclass[arguments]{aastex631}
\shorttitle{Decimetric type U solar radio bursts}
\shortauthors{Gao et al.}
\graphicspath{{./}{figures/}}
\begin{document}

\title{Decimetric type U solar radio bursts and associated EUV phenomena on 2011 February 9}

\author[0000-0001-6383-6751]{Guannan Gao}
\affiliation{Yunnan Observatories, Chinese Academy of Sciences, Kunming,  650216, P. R. China}
\affiliation{Center for Astronomical Mega-Science, Chinese Academy of Sciences,  Beijing, 100012, P. R. China}

\author[0000-0001-5188-0165]{Qiangwei Cai}
\affiliation{Institute of Space Physics, Luoyang Normal University, Luoyang, 471934,  P. R. China}

%\collaboration{6}{(AAS Journals Data Editors)}

\author{Shaojie Guo}
\affiliation{Yunnan Observatories, Chinese Academy of Sciences, Kunming,  650216, P. R. China}
\affiliation{Center for Astronomical Mega-Science, Chinese Academy of Sciences,  Beijing, 100012, P. R. China}
\affiliation{University of Chinese Academy of Sciences, Beijing, 100049, P. R. China}

\author{ Min Wang}
\affiliation{Yunnan Observatories, Chinese Academy of Sciences, Kunming,  650216, P. R. China}
\affiliation{Center for Astronomical Mega-Science, Chinese Academy of Sciences,  Beijing, 100012, P. R. China}



\begin{abstract}

A GOES M1.9 flare took place in active region AR 11153 on February 9,~2011. With the  resolution of 200~kHz and a time cadence of 80 ms, the reverse-drifting (RS) type III bursts, intermittent sequence of type U bursts, drifting pulsation structure (DPS), and fine structures were observed by the \textit{Yunnan Observatories Solar Radio Spectrometer} (YNSRS). Combined information revealed by the multi-wavelength data indicated that after the DPS which observed by YNSRS, the generation rate of type U bursts suddenly increased 5 times than before. In this event, the generation rate of type U bursts may depend on the magnetic reconnection rate. Our observations are consistent with previous numerical simulations results. After the first plasmoid produced (plasma instability occurred), the magnetic reconnection rate increased suddenly  5-8 times than before. 
Furthermore, after the DPS,  the frequency range of turnover frequency of type U bursts is obviously broadened 3 times than before, which indicates the fluctuation amplitude of the density in the loop-top.  Our observations also support the numerical simulations during the flare impulsive phase. The turbulence occurs at the top of the flare loop, the plasmoids can trap the non-thermal particles and cause the density fluctuation at the loop-top. The observations are generally consistent with the results of numerical simulations, helping us to better understand the characteristics of the whole physical process of eruption.

\end{abstract}



\section{Introduction} \label{sec:intro}

Magnetic reconnection is a common phenomenon in the solar atmosphere and it can occur in a slowly changing way, which may contribute to coronal heating \citep[e.g.,][]{pri99}, but more often happen as sudden violent processes such as flares, coronal mass ejections (CME), jet eruptions, etc \citep[e.g.,][]{lin00,cas12}. Magnetic reconnection leads to plasma heating and particle acceleration, which in turn, can efficiently contribute to chromospheric evaporation, thermal conduction fronts, and solar radio bursts during the solar eruption \citep[e.g.,][]{tian14,ni15,ni16,ni17,hong17,hong19,cai18,bi20}.%done done

The energetic electrons are accelerated by magnetic reconnection, the instability of an electron beam can generate Langmuir (plasma) waves at the local plasma frequency that can be converted into electromagnetic emission. Then the electron beam propagates in upward direction (anti-sunwards) along open magnetic field lines producing normal type III solar radio bursts with a negative frequency drift rate.
\cite{cai18} provided the long-desired direct evidence that electrons energized in magnetic reconnection regions to produce type III solar radio bursts. In contrast, the energetic electrons propagate in downward direction (sunwards) to form the reverse-drifting (RS) type III bursts with a positive frequency drift rate, as the energetic electrons move mirror in a closed magnetic field line and show first a negative drift rate and later a positive drift rate after they crossed the loop apex, which appears shaped as an inverted letter U, and thus are named type U bursts. In some cases, the energetic electrons are accelerated by magnetic reconnection and reach the mirror point near the opposite footpoint of the flare loop system, and then by the magnetic mirror effect, the energetic electrons moving along the loops and type U bursts are formed. A partial type U burst which does not show a fully developed downward branch is named a J burst \citep[e.g.,][]{asc93,wang01,ning00,ning03,kong16}. The physical relevance of type III and U bursts is the diagnostics whether the electron beams propagate along open or closed magnetic field lines \citep[e.g.,][]{asc09}.   %done done
 
The radiation mechanism of type III, U bursts, and their subtypes is usually considered as plasma radiation, which produces radiation at the plasma frequency (or second harmonic frequency),
and its square is directly proportional to the electron density. Therefore, the coronal electron density distribution can be diagnosed by the frequency distribution of type III, type U bursts, and their subtypes. Especially, the type III pair bursts occurring in flare impulsive phase have lower separate frequencies,  indicates the plasma density of the magnetic reconnection energy release region \citep{tan16a,tan16b}. The coronal altitude of the magnetic reconnection region can be estimated by the coronal density models \citep[e.g.,][]{asc95}. %done done

In the numerical simulations of the magnetic reconnection in current sheet, \cite{she11} found that the rate of magnetic reconnection, M$_{A}$, after the first plasmoid generation was about 5 to 8 times higher than before. Similar results were found in numerical simulations of \cite{ni18},  their results also demonstrated after appearing the plasmoid instabilities,  the reconnection rate sharply increased by a factor of about 5 times. This phenomenon of magnetic reconnection was not observed due to the limited resolution of observations.% done done

 In addition, the jets are usually considered to be the results of the plasma heated and accelerated by magnetic reconnection and moving along open or large scale magnetic field lines. Therefore, they often have straight or slightly curved structures. The observations also show that there are lots of fine structures and radio bursts in the process of jets \citep[e.g.,][]{she11,shen12,zhang17,shen18}.%done done..

\section{Observations and Results} \label{sec:Observations and Results}

 A GOES M1.9 class flare in soft X-rays occurred in active region AR 11153 on February 9, 2011. The onset, peak and end times are 01:23 UT, 01:28 UT and 01:31 UT, respectively. In the same time, several solar radio bursts and fine structures were observed by the\textit{Yunnan Observatories Solar Radio Spectrometer} \citep[YNSRS,][]{gao14a} and \textit{Hiraiso Radio Spectrograph} \citep[HiRAS,][]{kon95}.%done done
 
\begin{figure}[htbp]
\center{\includegraphics [scale=1.2]{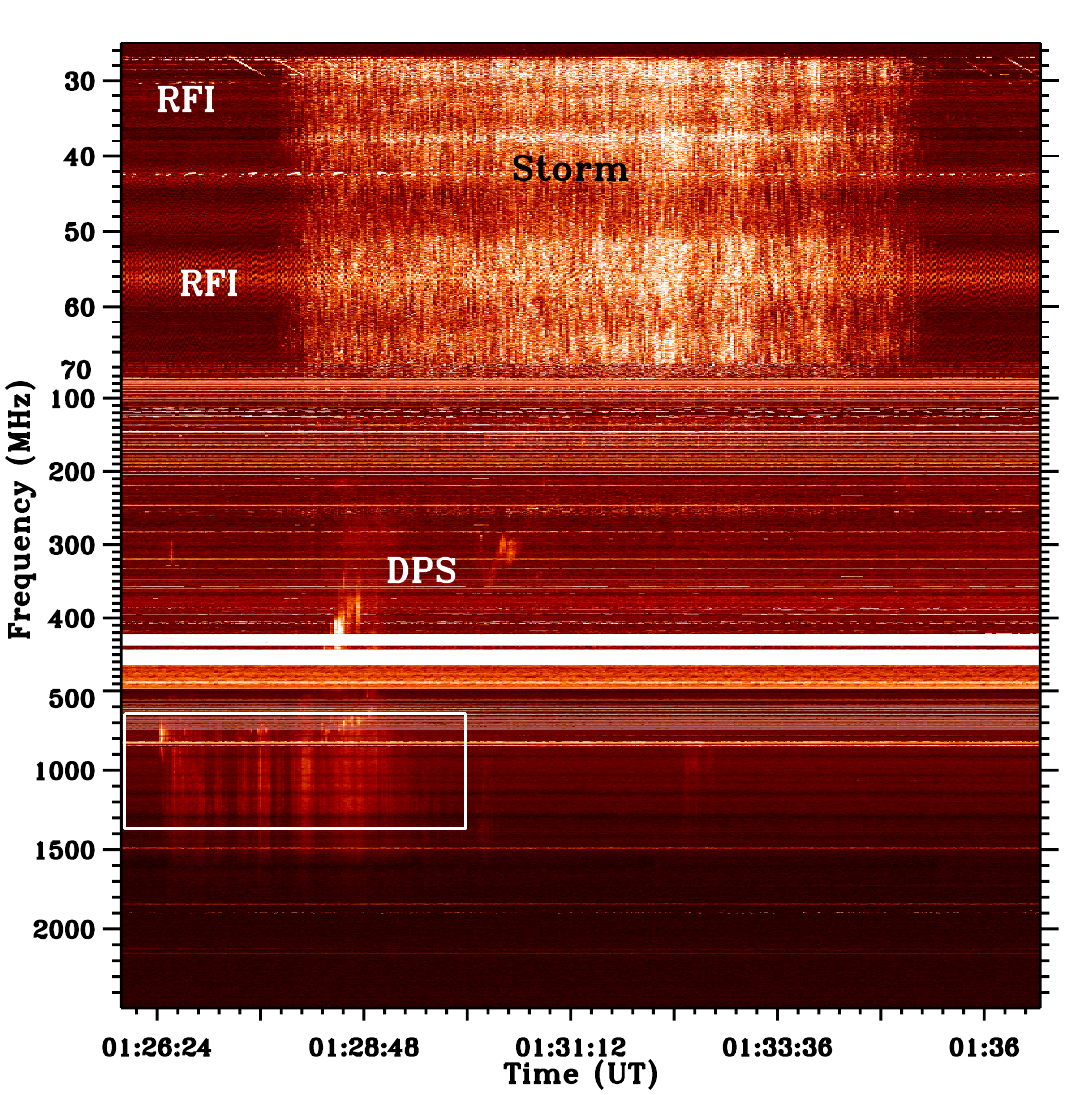}}
 \caption {\footnotesize Solar radio bursts observed by HiRAS-1, HiRAS-2, and HiRAS-3 from 25 to 2500 MHz on February 9, 2011. `RFI' means the radio frequency interferences, `Storm' and `DPS' structures are also displayed. The frequency range of YNSRS  data of Fig. 2a is indicated by the white box.}
 \label{fig:1}
\end{figure}

The solar radio bursts occurred between 01:26 and 01:36 UT on February 9, 2011 and were detected by HiRAS, which consists of three antennas, HiRAS-1, HiRAS-2, and HiRAS-3, which have the frequency ranges of 25$-$70, 70$-$500 and 500$-$2500 MHz with spectral resolution are 45 kHz,  430 kHz, and 2 MHz, respectively. Fig.~\ref{fig:1} displays the dynamic spectrum of this event obtained by HiRAS with a time cadence of 1 s. In the dynamic spectrum observed by HiRAS-3, between 01:26:24 and 01:30 UT, some type III bursts with narrow band embellish over a group of decimetric radio bursts (DCIM) occurred from 700 to 1500 MHz. Actually, these type III bursts observed by HiRAS are three groups of type U bursts observed by YNSRS with the higher time and frequency resolution (see Fig.~\ref{fig:2}a).  
Fig.~\ref{fig:1} also shows the drifting pulsation structure (DPS) drifting from 440 MHz at 01:28:21 UT to 340 MHz at 01:28:45 UT, as well as the two U bursts that are stringed together,
these type U bursts occurred from 220 to 300~MHz observed by HiRAS-2 between 01:30:15 and 01:30:42 UT.  The obvious radio storm, observed by HiRAS-1, occurred between 01:28:06 and 01:34:50 UT and the frequency range from 27 to 70 MHz.%done done

The dynamic spectrum in the white box of Fig.~\ref{fig:1} was also observed by YNSRS, which works in the frequency range from 625 to 1500 MHz with the spectral resolution of 200 kHz and a time cadence of 80 ms. The frequency range of 800$-$975 MHz has no data because of the notch filters that suppress serious interferences in this band (see \citealp{gao14a}). The dynamic spectrum of the event obtained by YNSRS is plotted in the Fig.~\ref{fig:2}a between 01:26 and 01:30 UT on February 9, 2011, which shows the reverse-drifting (RS) type III  bursts, type U bursts, fine structures, and DCIM which were also observed by HiRAS. We assigned the type U bursts into three groups with the obvious time interval of each group (Groups 1$-$3). These groups include 1 (01:26:42.280$-$01:26:42.920 UT), 6 (01:27:28$-$01:27:46 UT), and 16 (01:28:17$-$01:28:52 UT) type U bursts, respectively. We also note that it is hard to resolve type U bursts in Fig.~\ref{fig:1}, therefore, if there is a lack of high-resolution data, these type U bursts may be usually considered narrow-band type III bursts in previous observations. The continuous curve with yellow color in 
Fig.~\ref{fig:2}a is time derivative of GOES 1$-$8 \AA ~time series, which is used as a proxy for the missing HXR data because no RHESSI data at that time interval were available. Around the peak of the flare ($\sim$01:28 UT), the Group 3 type U bursts have the largest number and the fastest generation rate, meanwhile the rising and dramatic variability  happened in the time derivative curve of GOES 1$-$8 \AA ~time-series.  In addition, the red curve in 
Fig.~\ref{fig:2}a  is the \textit{Nobeyama Radio Polarimeters} \citep[NoRP,][]{nak94} flux plot at 1 GHz, which shows the maximum flux density of these radio bursts is about 300 s f u. %done  %done

In order to investigate the details of these radio bursts, we zoomed in each region of 
Fig.~\ref{fig:2}a, hence, Figs.~\ref{fig:2}b$-$\ref{fig:2}f are corresponding to type RS bursts, groups 1$-$ 3 type U bursts, respectively.
 Fig.~\ref{fig:2}b shows a group of  type RS bursts  between 01:26:24 and 01:26:38 UT. Combining the HiRAS-3 and YNSRS data revealed by Figs.~\ref{fig:1} and~\ref{fig:2}b, we found the type RS bursts drifting from 660 to 970 MHz with the drifting rate
of 660 $\pm 110$~MHz~s$^{-1}$. %done 这里写的是RS,fig b
 The first group (Group 1) has one type U burst (No.1 see Fig.~\ref{fig:2}b), which is after the type RS bursts and starts at 01:26:42.280 UT, as well as the time duration is about 640 ms.  Both the beginning and ending frequency are 800 MHz, as well as the turnover frequency is 766 MHz.%这里写的是figc图U burst done done
~Figs.~\ref{fig:2}c and \ref{fig:2}d display the second group (Group 2) of  the type U bursts (Nos.2$-$7), which occurred  between 01:27:28 and 01:27:46 UT. The Nos.2$-$4 and  6 have almost the same beginning, ending and turnover frequency, which are 800, 800 and 766 MHz, respectively. The turnover frequencies of Nos. 5 and 7 are lower than the others, which are about 740 and 700~MHz, respectively. 
Figs. \ref{fig:2}e and \ref{fig:2}f show the last group with 16 type U bursts (Nos.8$-$23) between 01:28:17$-$01:28:52 UT, which is the largest number of type U bursts in this event. Some enhancements between them are not included.  Hence, 16 is a lower limit, possibly more type U bursts occurred in this group. In addition, there are some fine structures (FSs) in HiRAS and YNSRS data, such as drifting pulsation structures (DPSs) (see Figs. \ref{fig:1} and \ref{fig:2}e).%\done

DPS is considered to be the radio signatures of  plasmoids forming and moving during magnetic reconnection in the impulsive phase of  the flare \citep[e.g.,][]{kli00,liu10,gao14b}. In the numerical simulations of magnetic reconnection, the emergence of plasmoids is considered to be a sign of plasma instability and turbulence \citep{shen11,ni15,ni16,ni17}.
Two DPSs in this event are displayed in Figs. \ref{fig:1} and \ref{fig:2}e, respectively. HiRAS observed the DPS with the center frequency drifting from 420 to 360 MHz, and the time interval is 01:28:21$-$01:28:45 UT, as well as the drifting rate is 
$-3$ $\pm~0.5$~MHz~s$^{-1}$ (marked with `DPS'  in  Fig. \ref{fig:1}). The other DPS after No.9 type U burst with the drifting rate of $-6.2 \pm 2.7 $~MHz~s$^{-1}$ observed by YNSRS occurred in the higher frequency, the center frequency range is about from 728 to 712~MHz, and the time interval is  about 01:28:23$-$01:28:25 UT  (marked with `DPS-H'  in  Fig.~\ref{fig:2}e).% 这里写的是两个DPS drifted from 740 to 700 MHz and the drifting rate is about $\sim15$~MHz~s$^{-1}$.,  
In addition,  after the DPS of Fig. \ref{fig:1}, the crowded two type U bursts occurred between 300 and 220 MHz observed by HiRAS-2 with the time interval of  01:30:18$-$01:30:36 UT, the turnover frequency is about 290 MHz. %this is N burst Done done

Figs.~\ref{fig:2}c$-$\ref{fig:2}f display some FSs in type U bursts. Fig.~\ref{fig:2}c shows the No.5 type U burst which is superimposed over the No.4. Both of them have almost the same turnover time ($\sim$01:27:36 UT) but the different turnover frequency, which are 760  and 740 MHz, respectively. We also note that Nos. 10$-$18 type U bursts occurred after the `DPS-H' observed by YNSRS (see Fig. \ref{fig:2}e and \ref{fig:2}f), and their turnover frequencies are not concentrated in one point, but distributed in a frequency range of $\sim$30 MHz 
 (such as No.10 type U burst in Fig. \ref{fig:2}f). In contrast, before the `DPS-H', the turnover frequencies of type U bursts Nos.1$-$9 are distributed in the range of $\sim$10 MHz (such as No. 9 type U burst in Fig. \ref{fig:2}e).%done done

\subsection{Radio Data Analysis} \label{subsec:Radio Data Analysis}
 The physical mechanism of  type III and U bursts is considered to be plasma radiation. The energetic particle motion in the coronal plasma with the velocity of 0.1$-$0.3~c causes the Langmuir instability and fluctuation, and then the corresponding energy is transferred to the electromagnetic wave with the same frequency at the local plasma frequency and/or second harmonic frequency. Therefore, the frequency $f_{obs}$ of the observed radio bursts is related to the plasma frequency $f_{p}$ and the electron density $n_{e}$  in the source region by $f_{obs}=s f_{p}, ~~f_{p}~[kHz] = ~8.98 \sqrt{n_{e}~[{cm}^{-3}]}$, and thus to the altitude of the source region, if a coronal density model, $n_{e}$=$n_{e}(h)$, is given. Here $s$ stands for the fundamental ($s=1$) and for the harmonic ($s=2$) band, respectively \citep{mcl85}. %done done

%In our calculations, the coronal electron density $n_{e}(h)=n_{0}g(h)$ is described by the empirical model of \citet{sit99}, such that
%\begin{equation}
%g(h)=a_{1}z^{2}(h)e^{a_{2}z(h)}[1+a_{3}z(h)+a_{4}z^{2}(h)+a_{5}z^{3}(h)],
%\end{equation}
%where   $z(h)= 1/(1+h),$ $ a_{1}= 0.001292$
   %$a_{2}= 4.8039,$  $a_{3}= 0.29696,$
   %$a_{4}= -7.1743,$ $a_{5}= 12.321,$
%with~$g(0)= 1$, and $n_{0}$ = 10$^{10}$~cm$^{-3}$ being the electron density at the base of the corona.%done

In this work, we use the electron density model of \citet{asc95}
for the flare region:
\begin{equation}
n_{e}(h)=\left\{
\begin{array}{l}
n_{1}(h/h_{1})^{-p}~~~~~~~~h<h_{1},\\
n_{Q}\exp(-h/\lambda)~~~~h>h_{1},\\
\end{array}
\right.
\end{equation}
where  $h_{1}$ is a transition height at which the way $n_{e}$ depending on $h$ changes. This model is constrained by the electron density $n_{Q}$ at the base of the quiet corona and the density scale height~$\lambda$.  It was developed for the region that has been affected by the heating and chromospheric evaporation in the flare process. A smooth transition between the two regimes is obtained by requiring that the function and its first derivative be continuous at the transition height $h=h_{1}$. These continuity conditions determine  $h_{1}$ and the density $n_{1}=n_{e}(h=h_{1})$:
\begin{equation}
% \begin{array}{l}
h_{1}=p\lambda,   ~~~~n_{1}=n_{Q}\exp(-p),
% \end{array}
\end{equation}
with~~$p=2.38$, ~$\lambda=6.9\times10^{9}$~$\textrm{cm}$, ~and $n_{Q}=4.6\times10^{8}$~$\textrm{cm}^{-3}$.

\begin{figure}[htbp]
\center{\includegraphics [scale=0.8]{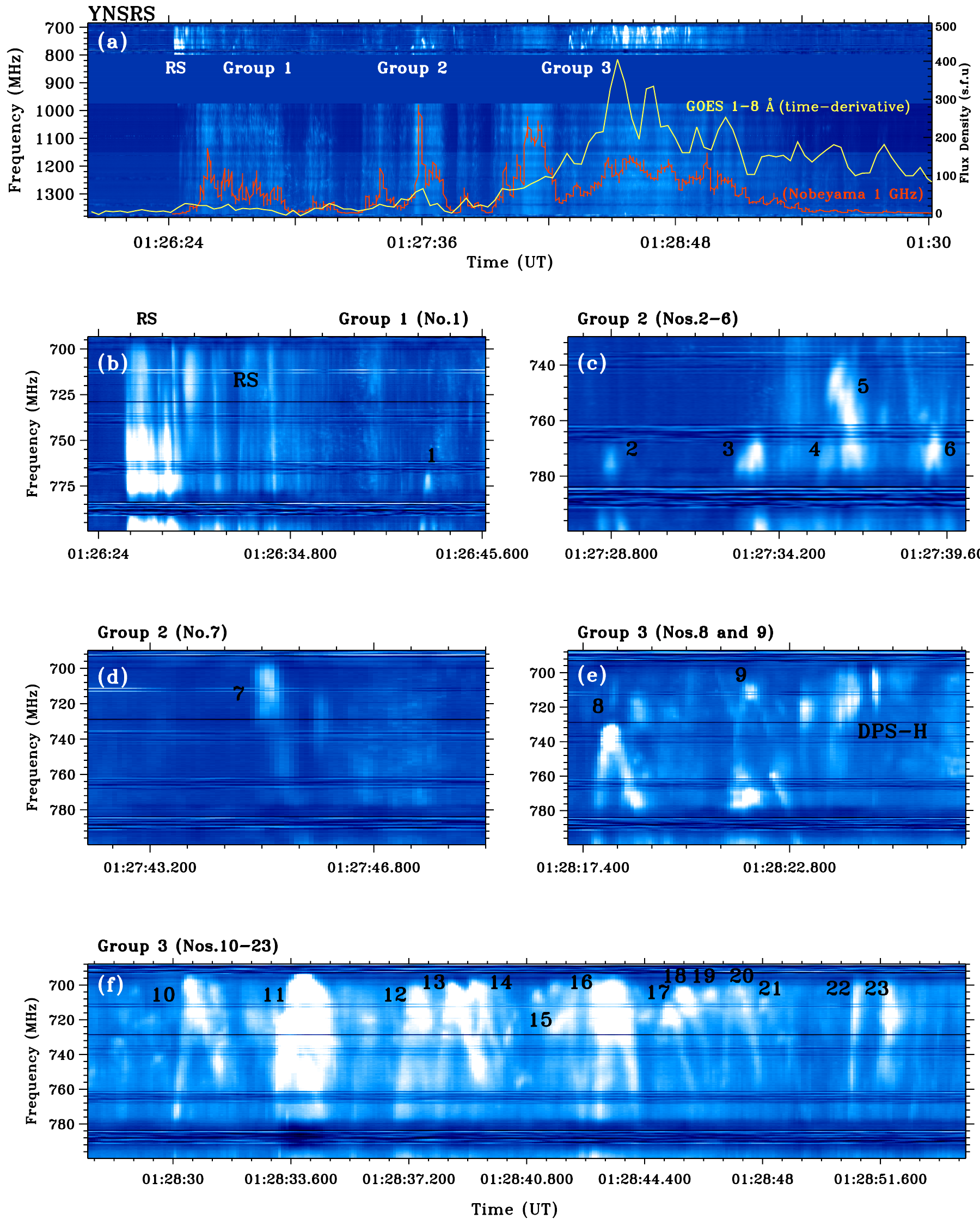}}
 \caption {\footnotesize (a) the dynamic spectrum of solar radio bursts observed  by YNSRS from 680 to 1380 MHz. `RS' and `Groups 1-3' type U bursts are marked. The continuous curve with yellow color is time derivative of GOES 1$-$8 \AA ~time-series, the red curve  is the Nobeyama Radio Polarimeters flux plot  at 1 GHz.
(b) the RS type III solar radio bursts occur at around 01:26:24$-$01:26:38 UT.
The first type U  burst  (No.1)  starts at around 01:26:42.280 UT, the during time is about 640 ms.
(c-d) the second group of type U bursts (Nos. 2$-$7) occur around 01:27:28$-$01:27:46 UT.
(e-f) the third group of type U bursts (Nos. 8$-$23) occur at around 01:28:17$-$01:28:52 UT, as well as the other drifting pulsation structure (marked with `DPS-H') showed in (e).  }
   \label{fig:2}
\end{figure}

 The type RS bursts were observed by YNSRS at the beginning of this event
 (see Figs.~\ref{fig:2}a and \ref{fig:2}b), they indicated the  electron beams  propagating in the sunward direction in contrast to the normal type III solar radio bursts. The starting frequency of the type RS bursts is $695\pm3$ MHz, according to the altitude of the region of particle accelerated by magnetic reconnection can be calculated by Equation (1), and is $20.6\pm0.1$~Mm (from the solar surface).%donedone

We investigated the radio data from YNSRS, the EUV data from \textit{Atmospheric Imaging Assembly} (AIA, \citealp{lem12}) and LOS magnetograms from \textit{Heliospheric and Magnetic Imager} (HMI,~\citealp{sch12}) on board the \textit{Solar Dynamics Observatory} (SDO). A newly emerging magnetic loop appeared below the existing magnetic loop at 01:25:45 UT in 131 \AA~band  (see Fig.~\ref{fig:5}a). At 01:26:21 UT, a disturbance of EUV was observed, which moved from one upper left footpoint of the original magnetic loop to another lower right footpoint  (see the red arrow in Fig.~\ref{fig:5}b). The cusp-shape structure formed at 01:26:26~UT  in 94~\AA~(see Fig.~\ref{fig:5}c), the loop top of the cusp-shape structure and the two footpoints brightened at the same time. The brightened areas are identified with red dashed box and orange arrows in Fig.~\ref{fig:5}c, respectively. Meanwhile,  the loop top of cusp-shape structure was  located in the junction of two opposite magnetic field (see the red dashed box in Figs.~\ref{fig:5}c and d, respectively). It indicates that the region of  magnetic reconnection may be locate in the current sheet above cusp-shape structure (see Figure~1 of Forbes \& Acton 1996). At the same time (01:26:24$-$01:26:38 UT), YNSRS observed the type RS bursts (see the right lower panel in Fig.\ref{fig:5}c and Fig.\ref{fig:2}b).  Combining with the EUV and radio data, we realize that the scenario may be the newly emerging magnetic loop reconnected with the original loop, forming a cusp-shape structure. Type RS bursts caused by the  electron beam accelerated by the magnetic reconnection and propagating from loop-top toward the footpoints (sunwards) in the cusp-shape structure. The accelerated electrons finally enter in the lower atmosphere and heat the lower atmosphere, resulting in the phenomenon of two footpoints brightening at the same time in 94~\AA~(see Fig.~\ref{fig:5}c).
%donedone

\begin{figure}[htbp]
\center{\includegraphics [scale=0.55]{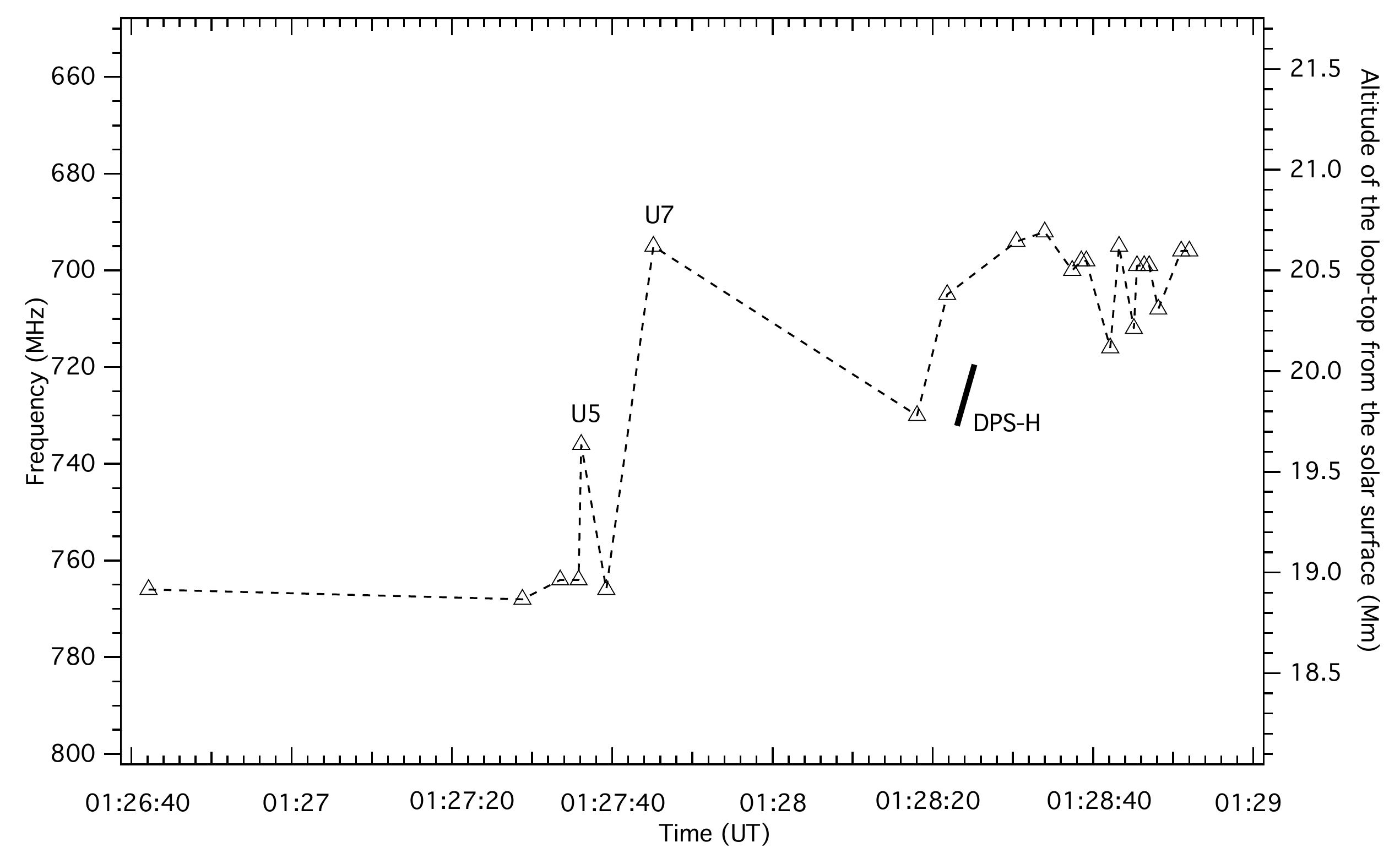}}
  
 \caption{\footnotesize The altitude and frequency of turnover frequencies of type U bursts, and DPS observed by YNSRS (`DPS-H').  Markers of triangle symbols represent turnover frequencies and associated altitudes of  type U bursts. The center frequency and altitude of `DPS-H' is indicated by a solid line.}
      \label{fig:4}
\end{figure}

\begin{figure}[htbp]
\center{\includegraphics [scale=0.7]{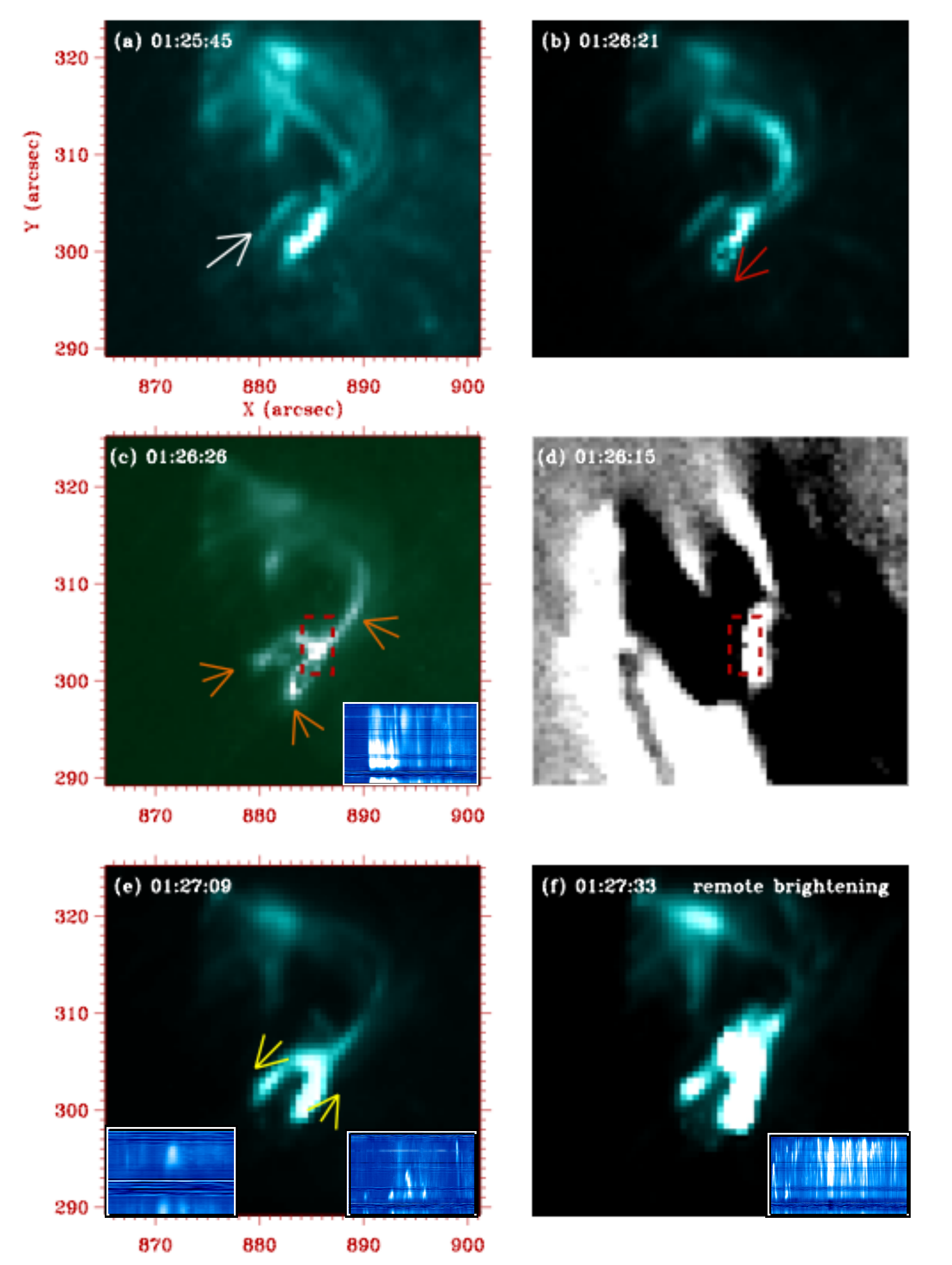}}
 \caption{\footnotesize (a) the white arrow points to the newly emerging loop in SDO/AIA 131 \AA . (b) the red arrow points to EUV inflows in the existing loop in 131 \AA . (c) the magnetic reconnection accelerates and heats particles in 94 \AA, the accompanying  RS type III  bursts in the same time are shown in the lower right panel. The orange arrows mark  the brightened areas of cusp-shape structure. The red dashed box indicates the region where the magnetic reconnection may occur. (d) shows  an HMI magnetogram, the red dashed box shows the same area as in (c),  it is also used to indicate that the possible area of magnetic reconnection with positive and negative polarities. (e) the phenomenon of chromosphere evaporation in 131 \AA~ is indicated by the yellow arrow and  the accompanying  type U bursts (Groups 1-2) display in the lower panels. (f) the cusp-shape structure continues to brighten and the remote brightening occur in
  131~\AA,  with the largest number type U bursts (Group 3) in the lower right panel.}
\label{fig:5}
\end{figure}

After the type RS bursts,  the type U bursts  (Nos.1$-$23) appeared, the turnover frequencies and  corresponding coronal altitudes deduced from  Equation (1) of type U bursts were displayed in Fig.~\ref{fig:4}. The turnover frequency of type U burst is the lowest frequency of each type U burst. In other words, the turnover frequency is a lower than the starting and ending frequency of each type U burst, respectively, which means that the trajectory of the electron beam first moves from the high-density coronal region to the low-density region, and then to the high-density region again.%donedone

Fig. \ref{fig:5}e shows that chromospheric evaporation phenomenon may be observed in the cusp-shape structure at 01:27:09~UT in 131 \AA. It is manifested as the brightening EUV propagating along the loop from the right footpoint to the left one (see the yellow arrows in Fig. \ref{fig:5}e). Meanwhile,  the Nos.1$-$7 type U bursts occurred (see lower right and left panels in Fig.~\ref{fig:5}e).  According to the sequence of various features shown 
in Figs. \ref{fig:5}a-\ref{fig:5}e, we acquire a scenario such that the electrons were accelerated by magnetic reconnection and the cusp-shape structure formed. The accelerating electrons propagate along the magnetic loop to the footpoints (sunwards) and produced type RS bursts, some energetic electrons were reflected by the magnetic mirror effect in one footpoint region and moved up along the loop to the loop-top (unti-sunwards), and then moved down to another footpoint (sunwards) producing the type U bursts. A part of the energetic electrons and heating conduction front reached the chromosphere (see Forbes \& Acton, 1996, Figure 1), the energetic electrons injected into the chromosphere and heated the chromospheric plasma, the chromospheric plasma propagating along the loop caused the brightening EUV  moving in the loop observed in 131 \AA~(see Fig. \ref{fig:5}e) with the velocity of about 360 km s$^{-1}$.%donedone（again）

Subsequently, at 01:27:33 UT, the remote brightening of the cusp-shape structure was observed in 131~\AA  (see Fig.~\ref{fig:5}f), and the whole cusp-shape structure rose up and brightened. The third group of type U bursts (Nos. 8$-$23, 01:28:17$-$01:28:52 UT), which is the type U bursts with the largest number and the fastest generation rate occurred at this event.  After the third group of type U bursts, a significant brightening in the whole EUV bands until the detector was saturated and could no longer distinguish the cusp-shape structure. At this time, the flare was at the maximum moment in soft X-ray observed by GOES.%done%done

We investigated the turnover frequencies of the type U bursts (Nos.1$-$23) and their corresponding coronal altitudes (see Fig. \ref{fig:4}). We notice that the turnover frequencies of all type U bursts occurred between 696$\pm$3 and 766$\pm$3~MHz, and the corresponding coronal altitudes were between 20.5$\pm$0.1 and 19.0$\pm$0.1 Mm (from the solar surface). Combining with the evolution of cusp-shape structure observed by SDO/AIA, the corresponding region of the turnover frequency should be the loop-top of cusp-shape structure. Therefore, the decreasing turnover frequency of type U bursts means  the rising altitude of the loop-top continuously, which indicated the expansion of the whole flare loop system. According to the turnover frequencies of type U bursts in Fig. \ref{fig:4}, we estimated the frequency drift rate is
$-~0.7~\pm$~0.1~MHz~s$^{-1}$ by the linear fitting, and the velocity of the whole flare loop system is 14.7~$\pm$~0.2~km~s$^{-1}$ (unti-sunwards) estimated by Equation (1). In addition, from the observation data of YNSRS and HiRAS, we have not found the second harmonic structure of type U bursts. Therefore, the calculation of coronal altitude is based on the fundamental frequency radiation, that is, $s$ = 1  in \textbf{$f_{obs}=s f_{p}$}. %done%done***

Fig. \ref{fig:4} shows the lowest turnover frequency of the type U bursts is 695$\pm$3~MHz, which corresponding to the  density of the loop-top is about 6.0($\pm$0.1$)\times 10^{9}$~cm$^{-3}$  by $f_{p}~[kHz] = ~8.98 \sqrt{n_{e}~[{cm}^{-3}]}$. The two bifurcations of the type U bursts extend to about 800$\pm$3 MHz, which corresponding to  the density of footpoints is  about 7.9($\pm$0.1$) \times 10^{9}$~cm$^{-3}$.%done done
 
\begin{figure}[htbp]
\center{\includegraphics [scale=0.66]{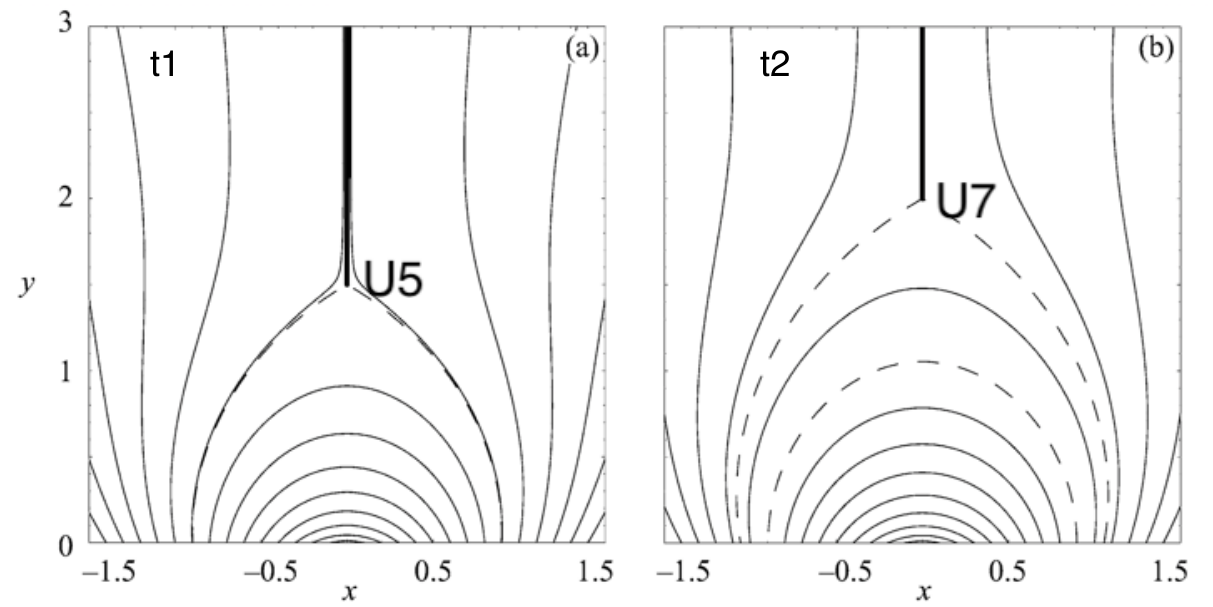}}
   \caption{\footnotesize Schematic demonstration of a new loop formed after magnetic reconnection and the whole flare loop system expanding; (a) the dashed curve indicates a new loop formed after magnetic reconnection. At this time, due to the magnetic mirror effect, the upward moving electron beam just moves along the newly formed loop, resulting in the No. 5 type U burst. (b)  another new  loop is formed at the magnetic reconnection point. Similar to the formation mechanism of No.5 type U burst, the new loops are constantly produced in the magnetic reconnection point, and the electron beam moves along the newly formed loops to form No.7 type U burst. }
      \label{fig:6}
\end{figure}

Both theories and observations assume that the flare loop system is expanding continuously during the solar eruption, which is manifested by the rising of flare loop system and the separation of footpoints in the corona. At the same time, a newly formed flare loop will be at a higher height of the corona than the others \citep[e.g.][]{sve87,lin95,for96,lin04,yan13,gao14b}. This behavior of the flare loop is not dependent on the plasma motion, but  the result of the region of magnetic reconnection moving to new magnetic field lines. We notice that the curve of the turnover frequency of type U bursts in Fig. \ref{fig:4} is very steep at U5 and U7, it indicates that the turnover frequencies of Nos. 5 and 7, type U bursts decrease significantly at that moments, also means the loop-tops rising significantly. In contrast, the turnover frequencies of the other type U bursts decrease relatively slowly, it indicates the slow rising of the loop-tops. The possible explanations of the Nos. 5 and 7 decreasing significantly is that the electron beams which reflected by the magnetic mirror effect move along the newly formed loops at a higher height in the magnetic reconnection region. Meanwhile, the turnover frequencies of most type U bursts decrease steadily, it  may demonstrate that the whole flare loop system is rising. We calculate the velocity  of the flare loop system rising is  14.7$\pm$ 0.2  km s$^{-1}$.
According to Figure 4 in Lin (2004), we sketch the formation process of U5 and U7 
(see Fig.\ref{fig:6}).%done%done

\begin{figure}[htbp]
\center{\includegraphics [scale=0.66]{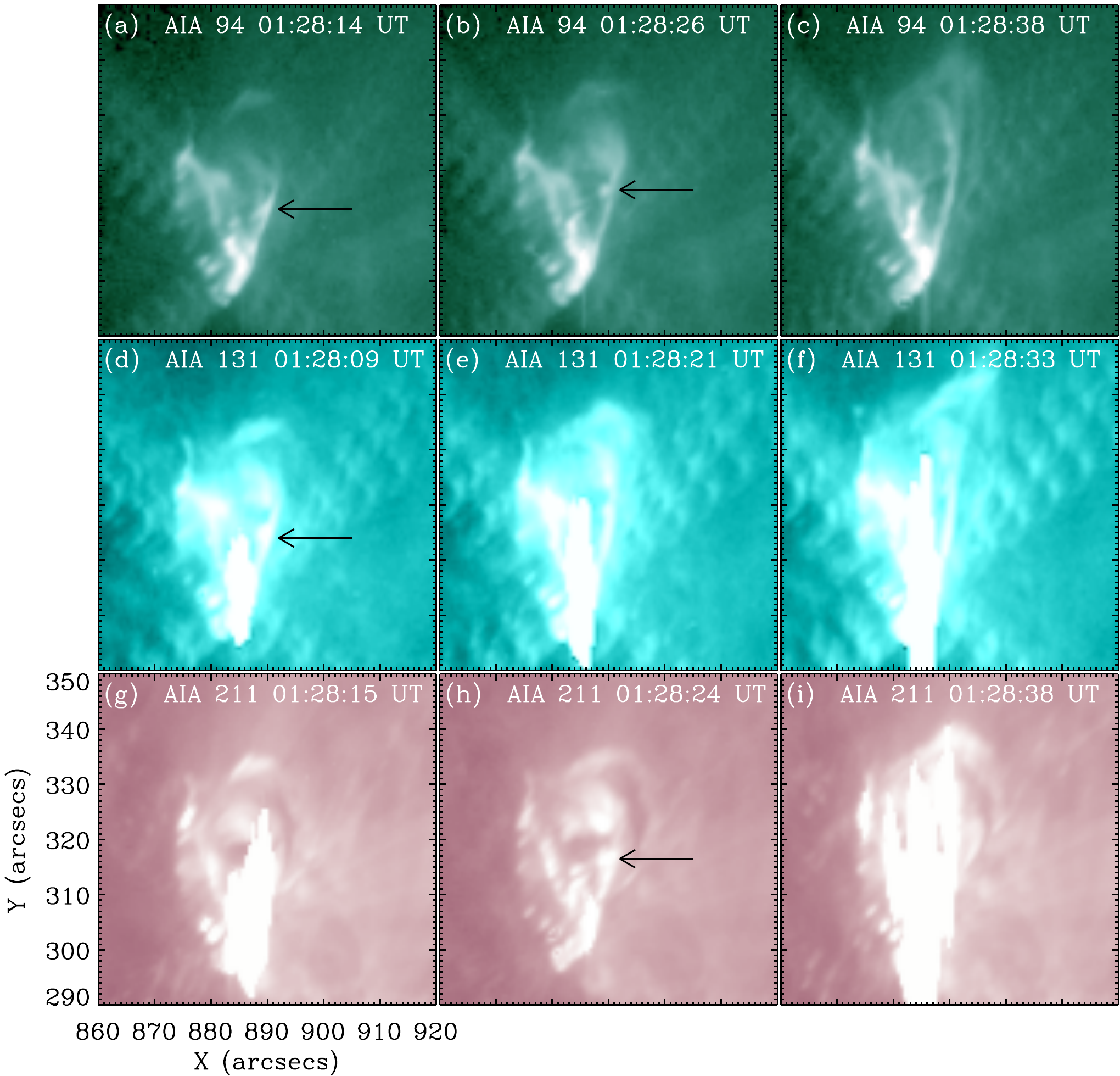}}
 \caption{\footnotesize During 01:28:14$-$01:28:38 UT, a plasmoid observed in AIA 94, 131, and 211 \AA~is indicated by black arrows. At the same time, the `DPS-H' is observed by YNSRS in Fig. 2e
 }
 \label{fig:7}
\end{figure}

It is worth noting that DPS in decimetric band is usually interpreted as radio signals of plasmoids formed during magnetic reconnection \citep[e.g.][]{kli00,bar08,gao14b,kar18}. Based on \citet{lin00} solar eruption model, \citet{shen11} studied the magnetic reconnection rate M$_{A}$  during magnetic reconnection through the numerical simulations, and found that  M$_{A}$ increased very slowly in the initial stage of magnetic reconnection, but increased significantly 5$-$8 times after the emergence of the first plasmoid or magnetic island \cite[see Figure 5 in][]{shen11}. Likewise, the numerical simulations of \citet{ni15,ni16,ni18} also demonstrated that the magnetic reconnection rate increased rapidly to 5 times after the emergence of plasma instability. In our work, the DPS occurred after the No. 9 type U burst observed by YNSRS at 01:28:23 UT and the frequency range is about 700$-$740~MHz see `DPS-H' in Fig.~\ref{fig:2}e). The center frequency and drifting direction of `DPS-H' are displayed with solid arrows in Fig.~\ref{fig:4}. According to the Equation (1), the velocity of the plasmoid is 137 $\pm$ 60 km s$^{-1}$ (anti-sunwards). Meanwhile, the structure of this plasmoid can also be clearly observed in SDO/AIA 94, 131, and 211 \AA~bands (see Fig.~\ref{fig:7}), especially in 94 \AA~(as shown by the arrows in Figs.~\ref{fig:7}a-\ref{fig:7}b). According to the SDO/AIA data, we obtain the velocity of the plasmoid is about 212 km s$^{-1}$ (anti-sunwards).  These two speeds are almost the same, the little difference of the plasmoid velocity calculated in EUV and radio data may be due to the coronal density model dependent or projection effect.%done notice the figures done

In this event, before the occurrence of `DPS-H' in Fig. \ref{fig:2}e, the generation rate of the type U bursts was about 1 every 10 seconds, and after the `DPS-H',  it increased to about 5 every 10 seconds. It manifested that after the emergence of the first plasmoid, the generation rate of type U bursts enhanced five times. Furthermore, Fig. \ref{fig:2}f displayed that some type U bursts after `DPS-H'  appeared intensively and caused some of them too close to resolve. Therefore, we assume that a factor of 5 times is just a lower limit. The type U bursts are generated by electron beams moving along the flare loop, so the generation rate of type U bursts may represent the generation rate of electron beams, and the magnetic reconnection rate could be described  by the generation rate of type U bursts. We suggest that this observed phenomenon may firstly support the results of  numerical simulations by~\citet{shen11,ni15,ni16,ni18}, in the process of magnetic reconnection, when the plasma instability occurs, the magnetic reconnection rate will increase significantly.%done%done

In addition, we also note that  type U bursts occurred before/after the `DPS-H' observed by YNSRS (see Figs. \ref{fig:2}e-\ref{fig:2}f), and their turnover frequencies are not concentrated in one point. Before the `DPS-H', the turnover frequency 
range of type U burst (such as No.9) is between 706 and 716 MHz,  as well as the frequency range is about 10 MHz (see Figs.~\ref{fig:2}e). However, after the  `DPS-H',  the frequency range of  turnover frequency is obviously broadened.  From Figs. \ref{fig:2}f, the frequency range of turnover frequency, such as No.10 in Fig.~\ref{fig:2}f, is between
696 and 726 MHz with the frequency range of 30 MHz. The duration of this state is 20 seconds, since then, the frequency range turns back to 10~MHz (see No.19 in Fig. \ref{fig:2}f).According to the sequence of features, the changes of turnover frequency range before/after the `DPS-H'  may  manifest the variation of the density at the loop-top,  we calculate the average density of the loop-top before the `DPS-H' (or the plasmoid generation), it is about  
6.3 $\times 10^9$~cm$^{-3}$, and the fluctuation amplitude of the density is $\pm~0.1$$\times 10^{9}$~cm$^{-3}$. On the other hand,
 after the plasmoid generation, the average  density of  the loop-top is about 6.3 $\times 10^{9}$ cm$^{-3}$, but the fluctuation amplitude of the density  is~$\pm0.3$$\times 10^{9}$ cm$^{-3}$.
It demonstrates  that the density at the loop-top has been in a significant fluctuation before and  after the `DPS-H'  (or the formation of the plasmoid), which is from 
6.3($\pm$0.1$)\times 10^{9}$ cm$^{-3}$ to 6.3($\pm$0.3$)\times10^{9}$ cm$^{-3}$. The observation results may support the numerical simulations of \citet{fang16}, they simulated the deposition of energy driven by magnetic reconnection in a flare loop during the flare impulsive phase, resulting in chromospheric evaporation between the two footpoints at a speed of hundreds of kilometers per second. The simulation results show that the turbulence phenomenon appears at the top of the flare loop and may be an effective non-thermal particle accelerator, and the magnetic island (or plasmoid) can capture non-thermal particles at the loop-top and cause the density fluctuation.%done done

%\begin{figure}[htbp]
%\center{\includegraphics [scale=0.8]{fig3new.pdf}}
%\caption {\footnotesize (a) the dynamic spectrum  observed by HiRAS-2 including the drifting pulsation structure (marked with  `DPS-L')  and  type N burst with markers 1 to 3 indicated stripes of word `N', respectively. (b$-$e) the dynamic spectrum of fine structures in the type U bursts observed by YNSRS including the other drifting pulsation structure (marked with `DPS-H') showed in (c). }
 %\label{fig:3}
%\end{figure}

\subsection{EUV Observations} \label{subsec:EUV Observations}

In order to better understand the physical process in this event,
we review this process from EUV observations. By analyzing the data from SDO/AIA  and HMI LOS magnetograms, the \textit{Solar Terrestrial Relations Observatory}
\citep[STEREO,][]{kai08}, we found this event including three procedures, EUV disturbances, magnetic reconnection, and jet eruptions.
\subsubsection{The first stage: EUV disturbances}
 % done
Before the radio bursts, the cusp-shape structure was not formed in SDO/AIA, and EUV disturbances occurred between 01:16:09 and 01:22:36 UT (see Fig. \ref{fig:8}). Four obvious EUV disturbances were observed in 94 and 171 \AA~bands, respectively, which manifested the rapid movement of the enhancement structures in the flare loop. According to the distribution of emission intensity in time sequence, we estimated that the velocity of the first and the second disturbances are 217.6 km s$^{-1}$ and 113.3 km s$^{-1}$, respectively. After 01:20:45 UT, the studied flare loop slightly rises up, which result in the unsuitable the speed of the third and the fourth disturbances. 

Figs. \ref{fig:8}a-\ref{fig:8}b show the first EUV disturbance observed in 94 and 171~\AA, respectively. Fig. \ref{fig:8}c shows the second EUV disturbance observed in 94 \AA, and Figs.~\ref{fig:8}e-\ref{fig:8}f display the third and fourth EUV disturbance, respectively.  We also note that all the EUV disturbances in this event come from the left upper footpoint of the original magnetic loop (the area framed with red dashed box in Fig. \ref{fig:8}c), the corresponding region is also drawn in the HMI magnetogram of Fig. \ref{fig:8}d. It manifests that  the left upper footpoint of the original magnetic loop is located at polarity inversion line. Therefore, the EUV disturbances propagating in the magnetic loop may come from the magnetic reconnection near the  left upper footpoint of loops. Furthermore, in this event, we have not observed the reflection phenomenon of EUV disturbance in the flare loops, which have been  reported by \citet{kum13} and \citet{kum15}.%done

\subsubsection{The second stage: magnetic reconnection}
The second process occurred between 01:25 and 01:28 UT (see Fig. \ref{fig:5}). A variety of radio bursts such as RS type III bursts, DPS  and the groups of type U bursts were observed by YNRSR and HiRAS. Around the peak of the flare ($\sim$01:28 UT), the third group of type U bursts has the largest number and the fastest generation rate, as well as the time derivative curve of GOES 1-8 \AA~soft X-ray flux also shows dramatic variability during this period (see the yellow curve of Fig. \ref{fig:2}a).%done

\subsubsection{The third stage: jet eruptions}
Figs. \ref{fig:9}a-\ref{fig:9}c display the evolutionary features of three jets 
observed by SDO/AIA 304~\AA~between 01:32:22 and 01:48:22 UT. The first and second jets are  straight jets (see Figs. \ref{fig:9}a and \ref{fig:9}b),  as well as the second jet has the obvious falling down (see the white arrow of Fig. \ref{fig:9}c).  The third jet propagates along the  magnetic loop rising and falling, together with two plasmoids are found in the third jet (see the Fig. \ref{fig:9}c).  In addition, because of the STEREO-A satellite position, the third jet appeared in the field of view (FOV) of STERE-A/EUVI A, and it showed the third jet with the same movement tendency as the SDO/AIA observed.
Furthermore, we  also investigate the velocity of each jet in Fig.~\ref{fig:10},
 Figs. \ref{fig:10}a and \ref{fig:10}b show the third jet including two parts, one part has the straight direction with the velocity of 403 km~s$^{-1}$ (anti-sunwards), and the other part propagates along the magnetic loop with the rising and falling speed of 106 km~s$^{-1}$. The dashed lines `CD' and `CE' in Fig.~\ref{fig:10}c are corresponding to the second and first jet motion trail, respectively. The velocity of the second jet is about 260 km~s$^{-1}$ and the falling part is in white dashed box in Fig.~\ref{fig:10}d.  Fig.~\ref{fig:10}e displays the first jet started at about 01:30 UT with velocity of  352 km s$^{-1}$, as well as no matter was found to fall back.%done 

\begin{figure}[htbp]
\center{\includegraphics [scale=0.68]{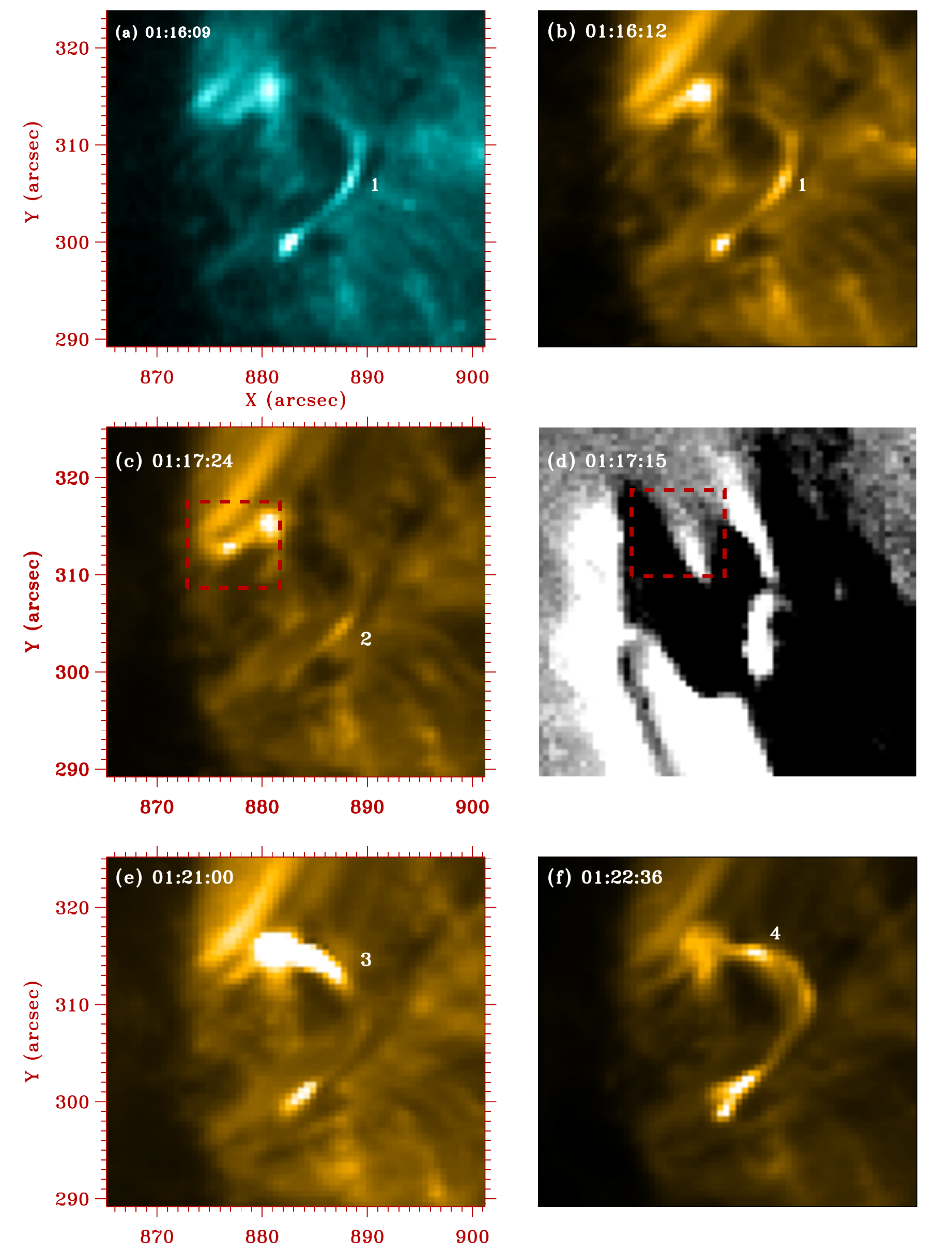}}
 \caption {\footnotesize The EUV disturbances are observed by AIA 94 and 171 \AA~between 01:16$-$01:23 UT. (a$-$b) the first EUV disturbance (No.1)  in both AIA 94 and 171 \AA,~respectively. (c) the second EUV disturbance (No.2)  in AIA 171 \AA. These disturbances propagate from the upper left footpoint to the lower right one. All of disturbances originate in the region of the red dashed box. (d) the red dashed box in HMI magnetogram also indicates that the  origin of the disturbances lies at the junction of positive and negative polarities. (e)$-$(f) Nos. 3 and 4 EUV disturbances in AIA 171 \AA.} 
\label{fig:8}
\end{figure}

\begin{figure}[htbp]
\center{\includegraphics [scale=0.68]{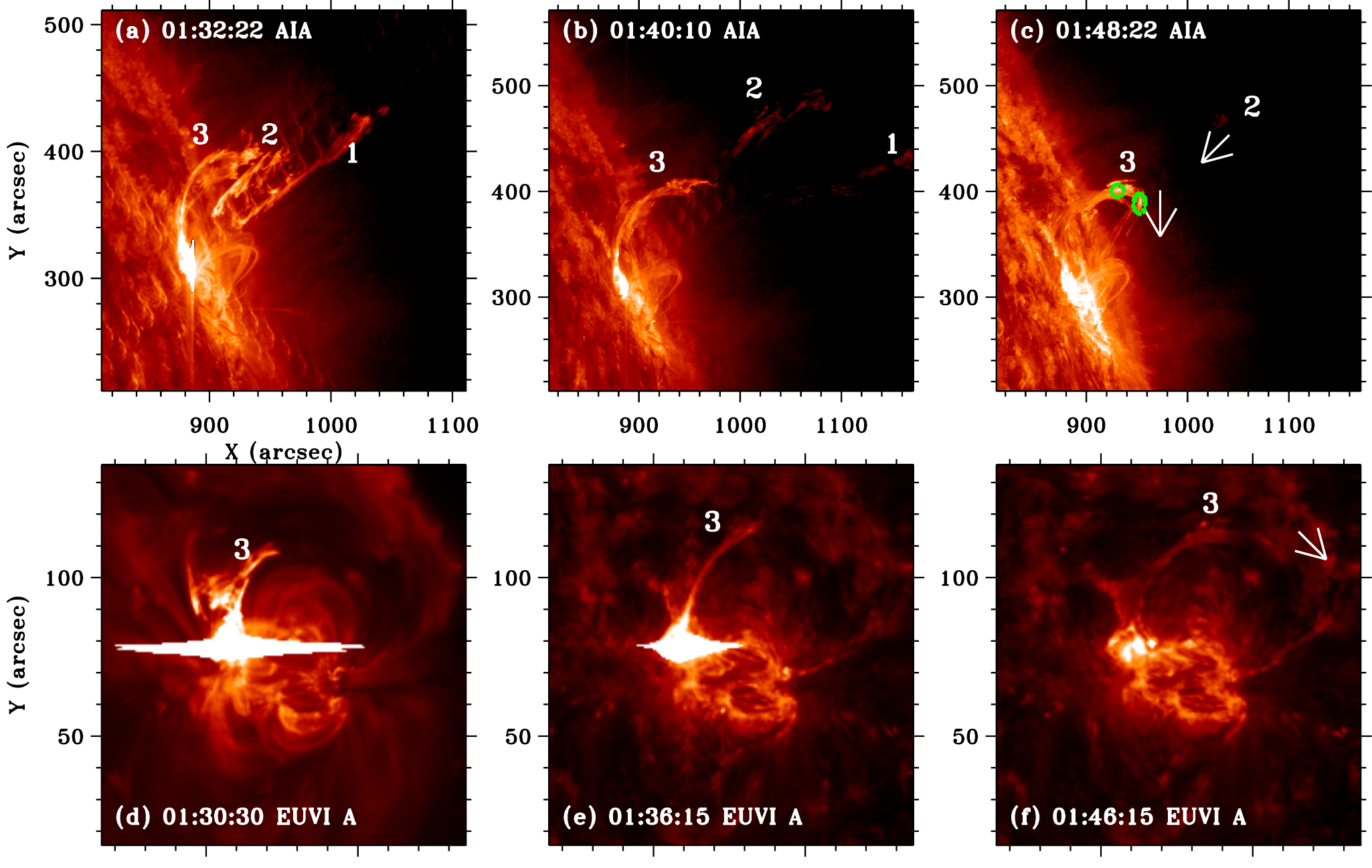}}
 \caption{\footnotesize After the type U  bursts, three jets are observed at 01:32:22$-$01:48:22 UT. (a$-$c) the white arrows indicate the direction of the jet,  the Nos. 3 and 2 jets have obviously falling back. Two plasmoids are marked by green circles in the No.3 jet. (d$-$f) the No.3 jet is also observed by STEREO-EUVI~A~304~\AA, the white arrow in (f) indicates the No.3 jet moving along a loop.}
 \label{fig:9}
\end{figure}

 \begin{figure}[htbp]
\center{\includegraphics [scale=0.7]{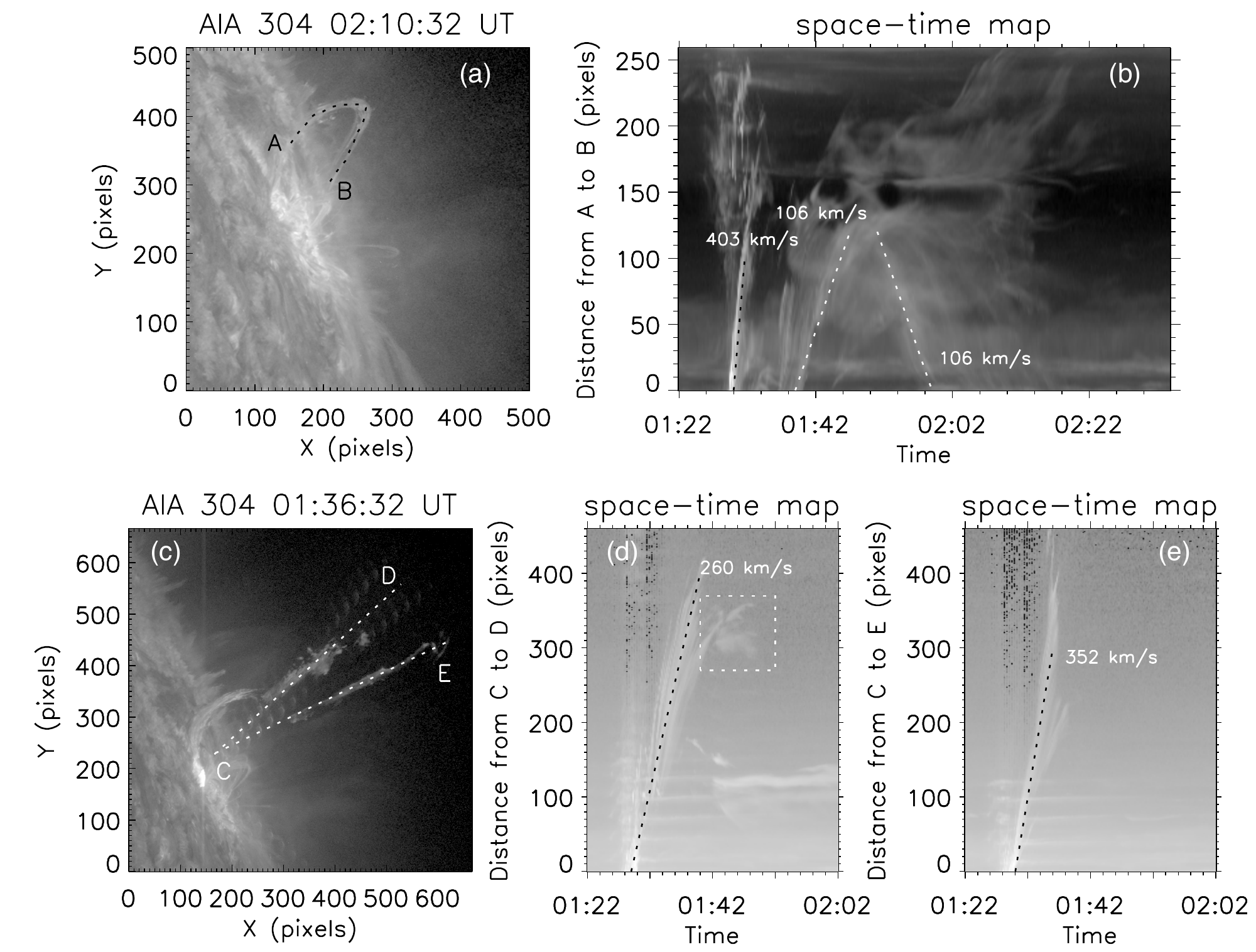}}
   \caption {\footnotesize{(a) time slices taken from the AIA 304 \AA~images at the position marked by the dashed curve AB, dash lines CD and CE in  (c),  which are corresponding to the jets  Nos.3$-$1 moving directions, respectively. The dashed lines in (b), (d) and (e) are used to derive the attached speeds, respectively.}}
      \label{fig:10}
\end{figure}

\section{Conclusions and discussions} \label{sec: Conclusions and discussions}

\begin{figure}
\centering 
\includegraphics[width=16.5cm,height=4cm]{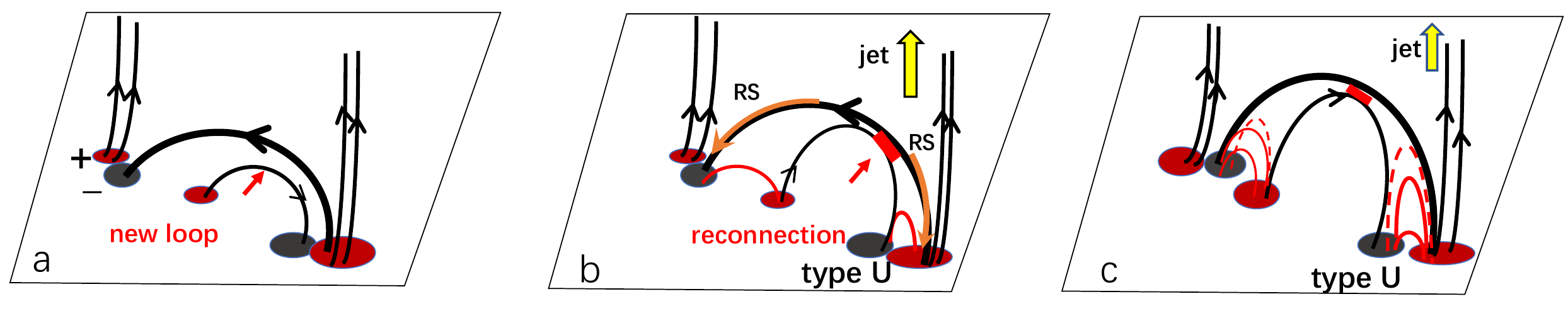}
\caption{Schematic illustration of the radio bursts and jet eruptions.  (a) the newly emerging loop  reconnects with the original loop above. (b) the electrons are accelerated by magnetic reconnection to high speed and propagating along the loops, as well as pouring into the footpoints (the RS type III solar radio bursts produced), partial energetic electrons propagating along the loop (red curves in b) by magnetic mirror effect,  and the type U bursts  are produced. (c) electrons propagate along the freshly loops to produced the Nos. 5 and 7 type U bursts  (red dashed curves in c). Magnetic reconnection also can lead to the formation of jets (yellow arrows in b and c)}
\label{fig:11}
\end{figure}

A GOES M1.9 class flare in soft X-rays occurred on February 9, 2011 including three stages: EUV disturbances, magnetic reconnection, and jet eruptions. In the first stage,  EUV disturbances propagate from the left upper corner footpoint of the original magnetic loop to the right lower corner footpoint (see Fig.~\ref{fig:8}). In the second stage, after the EUV disturbances, a newly emerging magnetic loop emerges around right lower corner footpoint
 (see Fig.~\ref{fig:5}a) and may be reconnect with the original magnetic loop to form cusp-shape structure (see Fig.~\ref{fig:5}c). Magnetic reconnection occurs in current sheet above the cusp-shape structure, and chromosphere evaporation may be found to occur between the two footpoints 
(see Fig.~\ref{fig:5}e). In the third stage, the magnetic reconnection releases energy and drives the jet eruptions. 
The scenario of this event is demonstrated in Fig.~\ref{fig:11}, in the magnetic reconnection, the high-resolution observation data of YNSRS show several type U bursts, which are hard to be resolved in HiRAS data at the same time. After No.9 type U burst, a DPS is recognized with a frequency drift rate of
 $-6.2 \pm 2.7 $~MHz~s$^{-1}$ (see `DPS-H' in Fig.~\ref{fig:2}e). According to the Equation (1), we calculate the velocity of the plasmoid represented by the DPS is is 137 $\pm$ 60 km s$^{-1}$ (anti-sunwards),  and the plasmoid is also observed by SDO/AA near the loop-top of cusp-shape structure with the velocity of 212 km s$^{-1}$ (anti-sunwards). %done
The difference of the plasmoid velocities measured in different wavelengths may be due to the projection effect and the uncertainty in the coronal density model. In addition, we measure the generation rates of type U bursts before and after the `DPS-H' as 1 per 10 seconds and 5 per 10 seconds, respectively. The observations are generally consistent with
the results of numerical simulations, helping us to better understand the characteristics of the magnetic reconnection.

We also note that the frequency range of turnover frequency of type U bursts before and after the `DPS-H'. Before the `DPS-H' (before the plasmoid formation), turnover frequency of type U burst covers a frequency range of 10 MHz. In contrary, after the `DPS-H'  (after the plasmoid formation), the frequency range of turnover frequency  is about 30 MHz, as well as  the latter is three times than the former one. This observations also support the \textbf{numerical simulations} of the energy deposition driven by magnetic reconnection during the flare impulsive phase, which produces chromospheric evaporation between the two footpoints, and the turbulence occurs at the top of the flare loop, and the plasmoids can capture the non-thermal particles and cause the density fluctuation of the loop-top.

By analyzing all the type U bursts in this event, we found that the turnover frequency of the whole type U bursts decreased slowly, except the turnover frequency of Nos. 5 and 7, decreased suddenly. This process can be explained by the physical process of the whole flare system is expanding, and the newly loop formed by magnetic reconnection is existing at the higher attitude than the previous loops.  The  energetic electron beam moving along the newly formed loop makes the turnover frequency of Nos. 5 and 7 decreased significantly.  We calculated the rising velocity of the whole flare loop system is 14.7$\pm$ 0.2  km s$^{-1}$ (unti-sunwards) estimated by Equation (1). We also studied the jets eruptions, after the magnetic reconnection, the velocity of jet eruptions  is about 200$-$400 km s$^{-1}$.

\begin{acknowledgments}
This work was supported by NSFC grants 11941003, 11403099, 11663007, 11703089, U1631130 and 11333007,  as well as  International Space Science Institute in Beijing (ISSI-BJ). Q. Cai was supported by grants from the Natural Science Foundation of Henan Province (212300410210). We acknowledge the use of  data from SDO, STEREO, and GOES statellites  and  Hiraiso radio spectrograph and Nobeyama Radio Polarimeters.
\end{acknowledgments}




\end{document}